\documentclass{jpsj2}
\def\d#1{\,{\rm d}#1}
\def\e{ {\rm e} }
\def\impsite{0}
\def\citeDOS{ {\cite{Oguri}} }

\def\pade{ Pad\'e }
\def\figTchivzero{\ref{VxTyTchiv0}(a) }
\def\figTchicvzero{\ref{VxTyTchicv0}(a) }
\def\figTchiv{\ref{VxTyTchi}(b) }
\def\figTchicv{\ref{VxTyTchic}(b) }

\title{Conductance through Quantum Dots Studied by Finite Temperature DMRG}
\author{Isao \textsc{Maruyama}$^{1}$, Naokazu \textsc{Shibata}$^{2}$ and Kazuo \textsc{Ueda}$^{1}$}
\inst{
$^{1}$Institute for Solid State Physics, University of Tokyo,
5-1-5 Kashiwa-no-ha, Kashiwa, Chiba 277-8581, Japan\\
$^{2}$Department of Basic Science, University of Tokyo
3-8-1 Komaba, Meguro, Tokyo 153-8902, Japan\\
}

\kword{quantum dot, Kondo effect, density renormalization group method, maximum entropy method,
\pade approximation}
\abst{
With the Finite temperature Density Matrix Renormalization Group method (F$T$-DMRG),
we depeloped a method to calculate thermo-dynamical quantities and the conductance of a quantum dot system.
Conductance is written by the local density of states on the dot.
The density of states is calculated with the numerical analytic continuation from the thermal Green's function which is obtained directly from the F$T$-DMRG.
Typical Kondo behaviors in the quantum dot system are observed conveniently by comparing the conductance with the magnetic and charge susceptibilities: 
Coulomb oscillation peaks and the unitarity limit.
We discuss advantage of this method compared with others.
}

\begin{document}
\maketitle

\section{Introduction}
Recently,
the Kondo effect is an attractive topic in quantum dots (QD).
At high temperatures, 
the conductance has peaks as a function of the gate voltage when the number of electrons in the dot changes,
i.e. coulomb oscillation peaks.
As a result of the Kondo effect,
the conductance becomes larger up to the unitarity limit ($2e^2 /h$) when the number of electrons is odd.~\cite{WielFFETK00}
Since a dot with an odd number of electrons plays a role of a magnetic impurity,
the conduction electrons in the source and drain leads tend to screen the local moment by the Kondo effect.
What is different from magnetic impurities is that the number of electrons is tunable from a magnetic region to a non-magnetic region by changing the gate voltage.
Additionally, in a lateral QD the strength of the tunnel coupling between the dot and the leads also changes by changing the gate voltage.
In this sense, not only temperature dependence
but also dependences of impurity parameters are important to understand the quantum transport through QDs.

For theoretical study of the conductance through QDs in a wide range of parameters,
the Finite Temperature Density Matrix Renormalization Group Method (F$T$-DMRG)~\cite{DMRG,Rommer} is used in this paper.
F$T$-DMRG has the advantage that we can re-use the result for the bulk part for different impurities or dots.
Strictly speaking, what we should do for various impurities is to average a quantum transfer matrix of the impurity site with the left and right eigen vectors which are obtained in the bulk-part calculations.
Additional advantage of the F$T$-DMRG is that not only static properties but also dynamical properties of the impurity site can be calculated.
However, F$T$-DMRG has not been applied to quantum dot systems, to the best of our knowledge.
One of the reasons may be the difficulty of the analytic continuation which will be discussed in the next paragraph.
For the application of the F$T$-DMRG and White's DMRG method~\cite{White92}, the system must be one-dimensional,
but this is not a serious problem for QD because we can use the model of QD as a impurity between semi-infinite tight-binding chains.

On the other hand, Wilson's Numerical Renormalization Group (NRG)~\cite{Wilson75,KrishnaWW80,IzumidaSS99} method is one of the most powerful methods to deal with the Kondo effect,
because the NRG is designed for lower temperature than the Kondo temperature which is exponentially small by focusing on the low energy excitations in logarithmic scale.
By the F$T$-DMRG it is difficult to reach that small temperature, while the high energy excitations are treated correctly.
This limitation of the lowest temperature originates from the numerical errors of the algorithm.
The numerical analytic continuation which is used to obtain Green's functions in frequency space from thermal Green's functions is the main source of the difficulty to decrease temperatures.
To avoid the difficulty we use relatively large tunnel coupling and coulomb energy in order to make the Kondo temperature high.
Fortunately the conductance which can be obtained from Green's function after the numerical analytic continuation has a better accuracy than the Green's function itself.

In the following sections,
we will show that in a certain parameter range conductance is obtained reliably from the F$T$-DMRG
and show the typical Kondo effect in QD: the Coulomb oscillation peaks and the unitarity limit.
In \S \ref{sym} we will discuss the parameter dependence of the Kondo temperatures in the symmetric case where the Kondo temperature has the lowest value as a function of the gate voltage.
In \S \ref{asym} we will show that there is the set of parameters which meets the condition and plot the typical Kondo effect in QD from the point of view not only of conductance but also of thermodynamic quantities.
In the next \S \ref{ham},
we start from definition of the model Hamiltonian and demonstration of how to calculate conductance by looking at the non-interacting case.

\section{Hamiltonian and the conductance in the non-interacting case \label{ham}}

\begin{figure}[tb]
%\resizebox{8cm}{!}{\includegraphics{fig/x-QD-vt-i.eps}}
\resizebox{8cm}{!}{\includegraphics{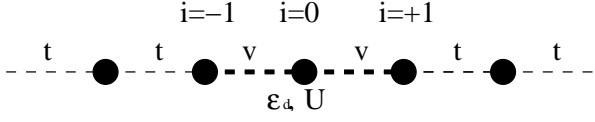}}
\caption{The tight binding chain with an dot $i=0$}
\label{x-QD-vt}
\end{figure}

F$T$-DMRG can be applied to various impurities not restricted to a single dot.
However in this paper we will concentrate on the Kondo effect in the single dot.
For this purpose, we will restrict ourselves to the simplest Hamiltonian,
which is a tight-binding chain with a dot which has a single local orbital,
\begin{eqnarray}
{\cal H}&:=& - \sum_{i\sigma} t_{i,i+1} 
( c^\dagger_{i\sigma} c_{i+1\sigma} + \mbox{h.c.} )
\nonumber \\
&&+\epsilon_d \sum_{\sigma} c^\dagger_{{\impsite}\sigma} c_{{\impsite}\sigma}
+Uc^\dagger_{{\impsite}\uparrow} c_{{\impsite}\uparrow}
c^\dagger_{{\impsite}\downarrow} c_{{\impsite}\downarrow}
\nonumber \\
&&+h \sum_{i\sigma} {\sigma  \over 2} c^\dagger_{i\sigma} c_{i\sigma}
\nonumber \\
&&+\mu \sum_{i\sigma} c^\dagger_{i\sigma} c_{i\sigma}
\label{H} \\
&t_{i,i+1}&:= 
\left\{
\begin{array}{cc}
v &\mbox{ ,when $i$ or $i+1$ is equal to $\impsite$} \\
t = 1 &\mbox{ ,others,} \\
\end{array}
\right.
\end{eqnarray}
where the site $i=0$ indicates the dot which has on-site Coulomb energy $U$ and $\epsilon_d$ corresponds to the gate voltage.
This Hamiltonian is sometimes called the Anderson-Wolff type impurity~\cite{AndersonWolff} (Fig.\ref{x-QD-vt}).
Magnetic field $h$ and Chemical potential $\mu$ are applied on the whole system.
We use $t$ as the units of energies.
This simplest model has been used by many authors for the study of QD,
because it includes the essential physics of the Kondo effect observed in experiments.

With this Hamiltonian, we will calculate thermodynamic quantities and a conductance in $h=\mu=0$ case.
Concerning the former, magnetic susceptibility ($\chi_{imp}=\partial \langle s^z_{i=0} \rangle/\partial h|_{h=0}$) and charge susceptibility ($\chi_{imp,c}=\partial \langle n_{i=0} \rangle/\partial \mu|_{\mu=0} $) are calculated to study behaviors of spin and charge fluctuations of the impurity site.
By this definition the units for the magnetic susceptibility are $(g \mu_{\rm B})^2$.
It should be emphasized that basic properties of these quantities are the same for the s-wave impurity-Anderson model.
This is easy to show, because this simplest Hamiltonian for QD can be mapped to two semi-infinite chains by using the liner combination such as ${1\over \sqrt{2}} (c_{i\sigma} \pm  c_{i \sigma})$; even-parity and odd-parity channels.
This situation is completely the same with the s-wave impurity-Anderson model attached to one-dimensional chain.
On the other hand,
transport properties of the two models are very different.
However, in the present paper we will restrict ourselves to the model given by eq.(~\ref{H}).

The latter, a conductance, can be calculated as follows.
Thermal Green's function in imaginary time is obtained from the F$T$-DMRG by calculating an expected value with respect to the eigen vectors having the largest eigen vectors.
In order to calculate the Green's function in frequency space from the thermal Green's function, we used the Maximum entropy method (MEM) \cite{MEM} and the \pade approximation~\cite{Pade} as numerical analytic continuation.
Since the MEM is basically a method to deal with statistical data, this method has been used with QMC to obtain the DOS of the QD~\citeDOS .
However, the MEM was also applied to the F$T$-DMRG and it has been shown that the MEM is more reliable than the \pade approximation\cite{Mutou}.
After the numerical analytic continuation, we can calculate the conductance from the following equation~\cite{Oguri,Sakai}, 
when the hopping terms from the impurity site to the left and right are the same.
\begin{eqnarray}
g&=&{2e^2 \over h} \int \d{\epsilon} 
{ 4\Delta_{\rm L} \Delta_{\rm R} \over \Delta_{\rm L} + \Delta_{\rm R}}
\left( -{\partial f\over \partial \epsilon}\right)
\left[ - {\rm Im} G_0(\epsilon) \right]
\label{eqg}
,
\end{eqnarray}
where $G_0(\epsilon)$ is the Green's function of the impurity site $i=0$ and $\Delta_{\rm L/R}$ is the local DOS of the next sites to the impurity site; $\Delta_{\rm L}=\Delta_{\rm R}=\Delta=v^2 \sqrt{4t^2 - \epsilon^2}/(2t^2)$ and $f$ is the Fermi distribution function.
It should be noted that $\Delta$ has some $\epsilon$ dependence in order to calculate the conductance at high temperatures which are comparable to the band width.
It should be also noted that at zero temperature the conductance follows a simple expression~\cite{NgL88} by the Freedel sum rule:
\begin{eqnarray}
g&=&{2e^2 \over h} \sin^2\left({\pi \langle n_{\rm d} \rangle \over 2}\right),
\label{eqg_sum}
\end{eqnarray}
where $\langle n_{\rm d} \rangle$ is the number of electrons at zero temperature and is defined as the difference between the total number of electrons $N_{\rm tot}$ and the system size $N$ i.e. $\langle n_{\rm d} \rangle=\langle N_{\rm tot} \rangle - (N-1)$.
Since $\langle n_{\rm d} \rangle$ is equal to $1$ in the symmetric case ($\epsilon_d =-U/2$),
we obtain quantized conductance $g=2e^2/h$ at zero temperature.

\begin{figure}[tb]
%\resizebox{8cm}{!}{\includegraphics{fig-new/conductanceU0.ver2.eps}}
\resizebox{8cm}{!}{\includegraphics{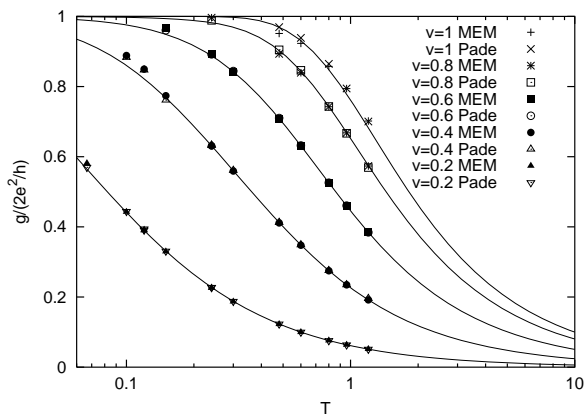}}
\caption{Conductance as a function of the temperature for $U=0$ at various $v$ calculated by the F$T$-DMRG with the finite algorithm.
The data are obtained with the MEM and the \pade from $G_0(\tau)$.
Trotter number $M \leq 200$ and the residual bases $m < 145$ are used.
The solid lines are analytic results:
$\rho(\epsilon)={1\over \pi v^2} {\sqrt{4-\epsilon^2} \over 4- v^{-2}(2-v^{-2})\epsilon^2}$.}
\label{conductanceU0}
\end{figure}

Before starting discussions on the Kondo effect of QD,
we first demonstrate reliability of the conductance obtained by the F$T$-DMRG with the numerical analytic continuation for the non-interacting case ($\epsilon_d=-U/2=0$).
In Fig.~\ref{conductanceU0},
we compare the results obtained by the \pade and the MEM with the exact ones (the solid lines in Fig.~\ref{conductanceU0}).
Conductances calculated with using these two methods show the tendency that errors become more significant closer to the unitarity limit: $g=2e^2/h$.
In fact, at lowest temperatures, some data are larger than the unitarity limit and these unphysical data are out of the range of the panel and not displayed.
In conclusion we may say that the numerical results of the conductance are reliable when the conductance is away from the unitarity limit,
except for the $v=1$ case where the local DOS at the impurity site becomes singular at the band edge.

\section{Kondo singlet and Unitarity limit for the symmetric case\label{sym}}

\begin{figure}[tb]
%\resizebox{7.5cm}{!}{\includegraphics{fig-new/krishna_maru.eps}}
\resizebox{7.5cm}{!}{\includegraphics{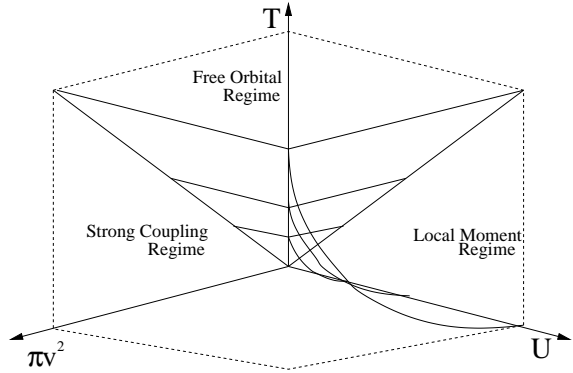}}
\caption{The schematic diagram in the symmetric Anderson model, reproduced from Fig.12 in \cite{KrishnaWW80}}
\label{krishna}
\end{figure}

\begin{figure}[tb]
%\resizebox{8cm}{!}{\includegraphics{fig-new/chiimpU2.eps}}
\resizebox{8cm}{!}{\includegraphics{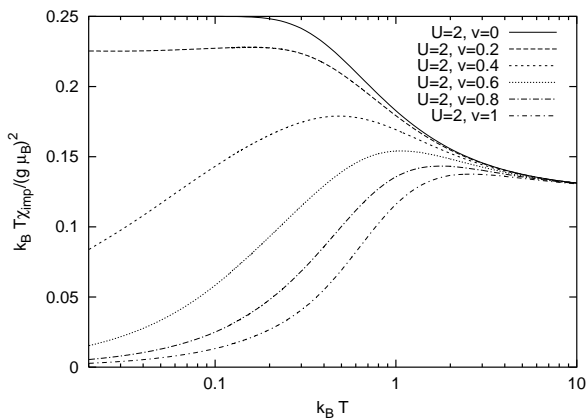}}
\caption{Susceptibilities $T\chi_{\rm imp}/(g\mu_{\rm B})^2$ as the function of temperature for $U=2$ and various $v$ calculated by the F$T$-DMRG with the infinite algorithm. The truncation error is $10^{-3} \sim 10^{-4}$, the maximum number of the residual bases $m$ is less than 145 and the Trotter number $M$ is less than 200.}
\label{chiimpU2}
\end{figure}

Since F$T$-DMRG is more reliable at higher temperatures,
we should arrange parameters in this Hamiltonian so that the Kondo temperature is sufficiently high.
Since the Kondo temperature as a function of the gate voltage has the lowest value in the symmetric case ($\epsilon_d=-U/2$) ~\cite{WielFFETK00},
the symmetric case is the most difficult case to see the Kondo effect.
In this section, therefore,
we restrict ourselves to the symmetric case and discuss the dependence on the other parameters, $U$ and $v$.
In the symmetric case,
the impurity susceptibility shows a Curie low ($T \chi \sim 1/4$ for 1/2 spin) in the temperature range where the local moment is well defined in the sense that the coupling to neighboring sites is negligible.
However,
due to the Kondo effect the local moment is screened ($\chi = $ Const.) at the low temperature limit:
the strong-coupling regime in terms of the scaling theory.
In contrast to the Kondo model or the s-d model,
the Anderson model has another high temperature limit ($T \chi \sim 1/8$ at $T \gg \pi v^2, U$): the free-orbital regime ~\cite{KrishnaWW80}.
The authors of \cite{KrishnaWW80} wrote a schematic diagram like Fig.~\ref{krishna}.
In this schematic diagram  we find that there is a direct cross-over from the free-orbital regime to the strong-coupling regime in the region $\pi v^2 > U$.
This strong-coupling regime is continuously connected to the non interacting limit $U=0$.
In fact, magnetic susceptibilities $T\chi_{\rm imp}$ for a typical case of $U=2$ (Fig.~\ref{chiimpU2}) obtained from the F$T$-DMRG show a good consistency with Fig.~\ref{krishna}.
Each lines start from the free-orbital regime at high temperatures.
Especially $v=1$ line shows a direct cross-over to the strong-coupling regime.
On the other hand, $v=0.2$ line shows clearly the local-moment regime,
but the strong-coupling regime cannot be seen
because the Kondo temperature $T_K$ is lower than 0.02.

\begin{figure}[tb]
%\resizebox{8cm}{!}{\includegraphics{fig-new/conductanceU2.ver2.eps}}
\resizebox{8cm}{!}{\includegraphics{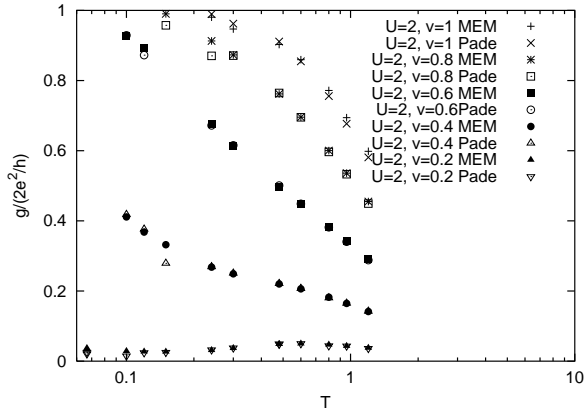}}
\caption{Conductance as a function of the temperature for $U=2$ with various $v$ calculated by the F$T$-DMRG. 
The other conditions are the same as in figure \ref{conductanceU0}.}
\label{conductanceU2}
\end{figure}

Similarly, conductance also shows a consistency with the schematic diagram.
The low (and high) temperature limit can be understood from eq.~\ref{eqg} (eq.~\ref{eqg_sum}) and the conductance $g$ is equal to $2e^2/h$ at zero temperature.
In fact, conductance for a typical case $U=2$ (Fig.~\ref{conductanceU2}) shows that each line starts from the free-orbital regime $g\sim 0$ at high temperatures.
Especially $v=1$ line shows a direct cross-over to the strong-coupling regime $g\sim 2e^2/h$.
On the other hand, $v=0.2$ line does not show clearly the local-moment regime 
in contrast to the magnetic susceptibility.
The local-moment regime in conductance seems to be difficult to distinguish from the free-orbital regime if we identify the local-moment regime simply by vanishingly small $g$.
This naive identification is consistent with the $v=0$ case where $g=0$ at any temperature.
However $v=0.2$ line in Fig.~\ref{conductanceU2} shows a slight peak at $T \sim 1$,
which may be on indication of the crossover between the free-orbital regime and the local-moment regime.
Strictly speaking, this broad hump comes from the double-peak structure of the local DOS at $\omega = \epsilon , \epsilon+U$ through the integration (eq.~\ref{eqg}).
Then, we find that the peak position of this conductance is around $T\sim U/2$ and it's height becomes zero at $v=0$.
It should be noted that the results based on \pade approximants show sometimes instabilities in Fig.~\ref{conductanceU2} (for example, $v=0.4$ and $T=0.15$),
because numerical errors are accidentally accumulated in the iteration process of the Thiele's reciprocal method~\cite{Pade}.
Because of the numerical analytic continuation,
the conductance is more difficult quantity than the magnetic susceptibility.
However, conductance obtained after integration in frequency space is fairly reliable except for some special cases as is seen in Fig.~\ref{conductanceU2}

\begin{figure}[tb]
%\resizebox{8cm}{!}{\includegraphics{fig-new/sz.vxUyT01.eps}}
\resizebox{8cm}{!}{\includegraphics{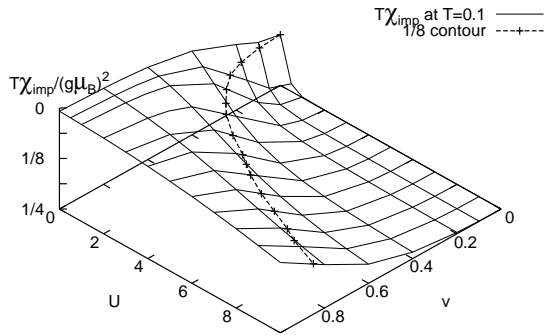}}
\caption{The magnetic susceptibility $T\chi_{\rm imp}$ as a function of $U$ and $v$ in the symmetric case at fixed $T=0.1$.
Three regimes are identified :the strong-coupling regime as $T\chi_{\rm imp} \sim 0$ , the local-moment regime as $T\chi_{\rm imp} \sim 1/4$ and the free-orbital regime around the origin, which is the region $\pi v^2 ,U < T$.
}
\label{sz.vxUyT01}
\end{figure}

\begin{figure}[tb]
%\resizebox{8cm}{!}{\includegraphics{fig-new/g-c.vxUy4f80.eps}}
\resizebox{8cm}{!}{\includegraphics{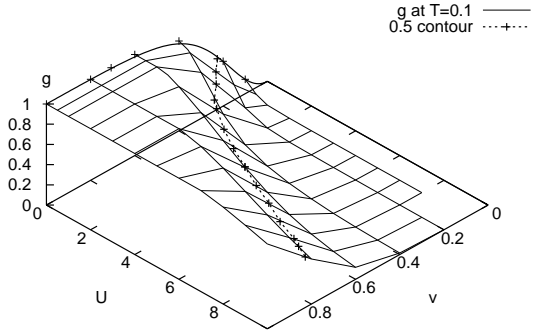}}
\caption{The conductance $g$ as a function of $U$ and $v$ in the symmetric case at fixed $T=0.1$.
The units of $g$ are $2e^2 / h$.
}
\label{g-c.vxUy4f80}
\end{figure}

From Fig.~\ref{chiimpU2} and ~\ref{conductanceU2} one can see that it is difficult to reach the strong coupling regime by the present method within the lowest temperatures when $v$ is small.
To determine the region where the strong coupling regime is reachable,
we fix the temperature ($T=0.1$) and plot the magnetic susceptibility (Fig.~\ref{sz.vxUyT01}) and the conductance (Fig.~\ref{g-c.vxUy4f80}) as a function of $v$ and $U$.
Note that the $z$-axis of Fig.~\ref{sz.vxUyT01} increases downward.
These two graphs show a similar behavior concerning the crossover from the local-moment regime to the strong-coupling regime.
The difference between the two figures is concerned with the free-orbital regime (around $\pi v^2 < T$ and $U < T$),
because the local-moment regime and the free-orbital regime are hard to distinguish for the conductance.
With the remark in mind, one can conclude that the cross-over behaviors from the local-moment regime to the strong-coupling regime are consistent between the two quantities.


\begin{figure}[tb]
%\resizebox{8cm}{!}{\includegraphics{fig-new/margeTk/modify4.ver4.eps}}
\resizebox{8cm}{!}{\includegraphics{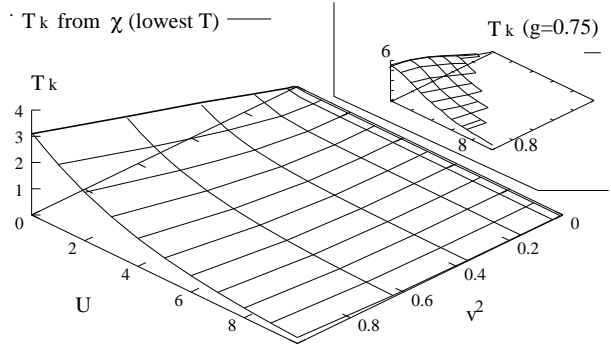}}
\caption{The Kondo temperature $T_{\rm K}$ estimated from the susceptibility $\chi_{\rm imp}$ at the lowest temperature $T=0.01$.
Inset: The Kondo temperature defined as $T_{\rm K}\mid_{g=0.75}={\pi^2 \over 2}T\mid_{g=0.75}$, 
where $T\mid_{g=0.75}$ is defined as the temperature where the conductance $g$ becomes $0.75 (2e^2/h)$. 
This estimated Kondo temperature is plotted only at high temperature region, because data of the conductance are available only $T\geq 0.1$ .}
\label{Tk}
\end{figure}

Finally, we try to estimate the cross-over temperature which is nothing but the Kondo temperature.
At zero temperature, all parameter region goes to the strong-coupling regime ($T\chi_{\rm imp}=0$  and $g=2e^2/h$) under the symmetric condition.
Physical aspects of the strong-coupling regime appear in the value of $\chi_{\rm imp}$ itself at zero temperature and in the way how $g$ approaches to the unitarity limit.
In $U\rightarrow \infty$ limit, the susceptibility at zero temperature is identified as $\chi_{\rm imp}=1/ (4 T_{\rm K})$, where $T_{\rm K}$ means the Kondo temperature.
This Kondo temperature also determines the temperature dependence of conductance: $g=(2e^2 /h)\left(1- {\pi^4 \over 16} ({T\over T_{\rm K}})^2+ O(T^4)\right)$.
The latter equation comes from the temperature dependence of the Kondo resonant peak in the DOS.
Figure ~\ref{Tk} shows the Kondo temperature obtained from $\chi_{\rm imp}$ and also from $g$ (Inset) by using the definitions of the large $U$ limit.
Since numerical data of $g$ do not have enough accuracy near the unitarity limit,
it is hard to estimate the coefficient of $T^2$ in $g$.
Therefore, we determine $T_{\rm K}$ simply from the temperature $T$ where $g=0.75$ with that equation.
$T_{\rm K}$ thus estimated from the conductance is only a rough estimation.
However, the two figures are consistent with each other.
Strictly speaking, these formulas which are used to define $T_{\rm K}$ are exact in the $U\rightarrow \infty$ limit where the Anderson model is mapped to the s-d model.
We consider that $T_{\rm K}$ shown in Fig.~\ref{Tk} qualitatively expresses the cross-over temperature to the strong-coupling regime.
In this sense,
we may consider that qualitative figure of Fig.~\ref{krishna} are reproduced numerically by Fig.~\ref{Tk}.

As a conclusion of this section,
we have observed behaviors in the strong-coupling regime both in magnetic susceptibility and conductance.
Concerning the latter, there is a parameter range where the conductance reaches the unitarity limit within the lowest temperature determined by the stability of the numerical analytic continuation.
It is concluded that the strong-coupling regime is easier to see when $v^2/U$ is large.
However,
to reproduce the typical Kondo behaviors which occur in the local-moment regime,
another limitation ($v^2/U$ is small) is needed.
In the next section, we will show that there is a region of parameters which satisfy the both limitations.
By choosing a parameter in the region, 
one can observe that the conductance changes from the Coulomb oscillation regime to the unitarity limit.

\section{Coulomb oscillation for the asymmetric case \label{asym}}
\begin{figure}[tb]
%\resizebox{8cm}{!}{\includegraphics{fig-new/conductanceasymv07U4pade4.eps}}
\resizebox{8cm}{!}{\includegraphics{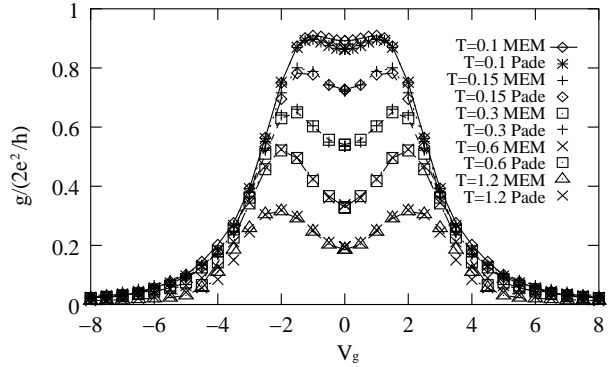}}
\caption{Conductances as a function of the gate voltage $V_{\rm g}$ for $v=1/\sqrt{2},U=4$ at various temperatures calculated by the F$T$-DMRG with the MEM and the \pade. 
 %{\color{red} The line labeled with `sum rule' indicates the conductance obtained with \ref{eqg_sum} in which $\langle n_d \rangle |_{T=0.04}$ are used.}
The other conditions for the numerical calculation are the same as in figure \ref{conductanceU0}. 
Since $\epsilon_d= -U/2 + V_{\rm g}$, $V_{\rm g}=0$ means the symmetric condition and $V_{\rm g}=2$ means $\epsilon_d=0$.
}
\label{conductanceaU4}
\end{figure}

In QD, the number of electrons is tunable by changing the gate voltage which corresponds to changing $V_{\rm g}$ in the asymmetric case ($\epsilon_d= -U/2 + V_{\rm g}$).
We stress again that F$T$-DMRG has the advantage to deal with the cases of different gate voltages at once.
As an important character of the asymmetric case, Coulomb oscillation peaks appear in the conductance as a function of the gate voltage,
because the conductance becomes non-zero only when the $n_{\rm d}=0$ and $n_{\rm d}=1$ configurations,
or the $n_{\rm d}=1$ and $n_{\rm d}=2$ configurations,
are degenerated , so-called the valence-fluctuation regime.
On the other hand, it is easy to understand the behavior of a conductance at zero temperature,
because the conductance is determined by $n_{\rm d}$ (eq.~\ref{eqg_sum}).
Especially in the case of $\langle n_{\rm d} \rangle\sim 1$, the conductance reaches the unitarity limit.
As we mentioned in the previous section,
there is numerical difficulty to see the Coulomb oscillation and the unitarity limit with the same $U$.
Since the Coulomb oscillation peak comes from the double peaks of local DOS at $\omega = \epsilon_d , \epsilon_d+U$,
the condition $U \gg \pi v^2$ is required.
We see clear coulomb oscillation peaks by putting $v^2/U$ smaller,
but it makes difficult to see the strong-coupling regime within $T > 0.1$.
Then, to compromise the conflicting requirements,
we take $U=4$ and $v=1/\sqrt{2}\sim 0.7$,
where $U$ is slightly larger than $\pi v^2 = \pi /2$.
The conductance for this set of parameters is shown in Fig.~\ref{conductanceaU4}.
In fact, Fig.~\ref{conductanceaU4} shows that the coulomb oscillation peaks at $V_{\rm g}=\pm 2$ are overlapping.
However, this figure shows the typical Kondo effect in the QD:
Coulomb oscillation peaks gradually move towards $V_{\rm g}=0$ at low temperatures and fill the valley up to the unitarity limit.

%{\color{red}
%Additionaly we plot the conductance caclucated by the sum rule, eq.(\ref{eqg_sum}) in Fig.~\ref{conductanceaU4}.
%$\langle n_d \rangle$ in eq.(\ref{eqg_sum}) is obtained from the F$T$-DMRG at $T=0.04$ and is exact at the symmetric point $V_{\rm g}=0$ where $\langle n_d \rangle$ = 1 at any temperature.
%Because of the property that $\langle n_d \rangle$ is exact at the symmetric point,
%the conductance caclucated by eq.(\ref{eqg_sum}) and the conductance at zero temperature are close.
%The reason why the structure around $V_{\rm g}=0$ is not plateau is that {\bf $v$ is large mainly}.
%}

Again, it is instructive to compare conductance and magnetic susceptibility.
To make the situation simple, let us start with the atomic limit ($v=0$).
For the magnetic susceptibility in the atomic limit, 
there are several fixed points~\cite{KrishnaWW80}:
the free-orbital fixed point ($T\chi_{\rm imp}= 1/8 $),
the valence-fluctuation fixed point ($T\chi_{\rm imp}= 1/6 $),
the local-moment fixed point ($T\chi_{\rm imp}= 1/4 $)
and the frozen-impurity fixed point  ($T\chi_{\rm imp}= 0 $).
It is easy to calculate the susceptibility in the atomic limit:
$T\chi_{\rm imp} $=$ \e^{{\epsilon_d+U\over T}} $
$\left\{2\left(1+2\e^{{\epsilon_d+U\over T}}+\e^{{2\epsilon_d+U\over T}}\right)\right\}^{-1}$.
Actually, we can see the above mentioned regimes in the plot of this function for $U=4$ as shown in Fig.~\figTchivzero.
Figure ~\figTchiv is $T\chi_{\rm imp}$ obtained from the F$T$-DMRG for $U=4$ and $v={1\over \sqrt{2}}$.
One can see that a developed moment in the local-moment regime,
$|V_{\rm g}|<U/2$ and $T<U$,
is very suppressed by the effect of $v$.
The valence fluctuation regime,
which is characteristic to the asymmetric case and is expected to correspond to the Coulomb oscillation,
is not clearly seen due to relatively large value of $v$.
The valence fluctuation regime may be characterized by enhanced charge fluctuations.
Therefore, we plot the charge susceptibility in Fig.~\figTchicv.

\begin{figure}[tb]
%\resizebox{8cm}{!}{\includegraphics{fig-renew-win/sz2-eps.eps}}
\resizebox{8cm}{!}{\includegraphics{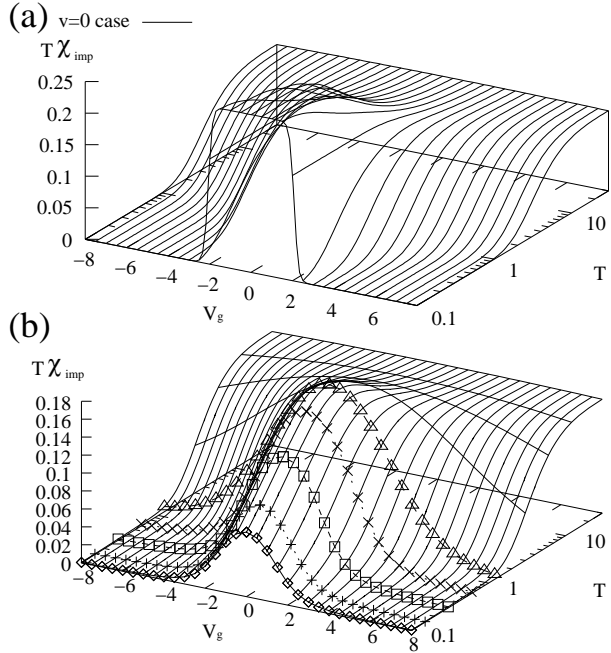}}
\caption{(a) Three dimensional plot of a spin susceptibility $T\chi_{\rm imp}$ as a function of $T$ and $V_{\rm g}$ for $U=4$ in the atomic limit $v=0$.
$T\chi_{\rm imp} = {\e^{{\epsilon_d+U\over T}}\over 2(1+2\e^{{\epsilon_d+U\over T}}+\e^{{2\epsilon_d+U\over T}})}$. 
(b) 
Three dimensional plot of a spin susceptibility $T\chi_{\rm imp}$ as a function of $T$ and $V_{\rm g}$ for $v=1/\sqrt{2}$. 
Data are plotted at $T=0.1, 0.15, 0.3, 0.6, 1.2, 3, 6, 12, 30$.}
\label{VxTyTchiv0}
\label{VxTyTchi}
\end{figure}

To understand the results we plot in Fig.~\figTchicvzero the charge susceptibility $T \chi_{\rm imp,c}$ in the atomic limit $v=0$ for $U=4$.
In the $v=0$ case, it is straightforward to obtain:
$T\chi_{\rm imp,c}$ = $2 \e^{{\epsilon_d+U\over T}} $
$\left(1+2 \e^{{\epsilon_d\over T}}+\e^{{2\epsilon_d+U\over T}}\right)$ 
$ \left(1+2\e^{{\epsilon_d+U\over T}}+\e^{{2\epsilon_d+U\over T}}\right)^{-2}$.
The value of $T\chi_{\rm imp,c}$ is $1/2$ in the free-orbital regime and $1/4$ in the valence fluctuation regime and zero in the other regimes.
Then we find that sharply defined valence fluctuation regime at $|V_{\rm g}|\sim U/2$, Fig.~\figTchicvzero, is suppressed by the finite $v$ ,Fig.~\figTchicv.
This behavior is the same as the magnetic susceptibility.
However, there remain small structures which reflect the suppressed valence fluctuation regime and are thought to correspond to the Coulomb oscillation peaks.

\begin{figure}[tb]
%\resizebox{8cm}{!}{\includegraphics{fig-renew-win/n2-eps.eps}}
\resizebox{8cm}{!}{\includegraphics{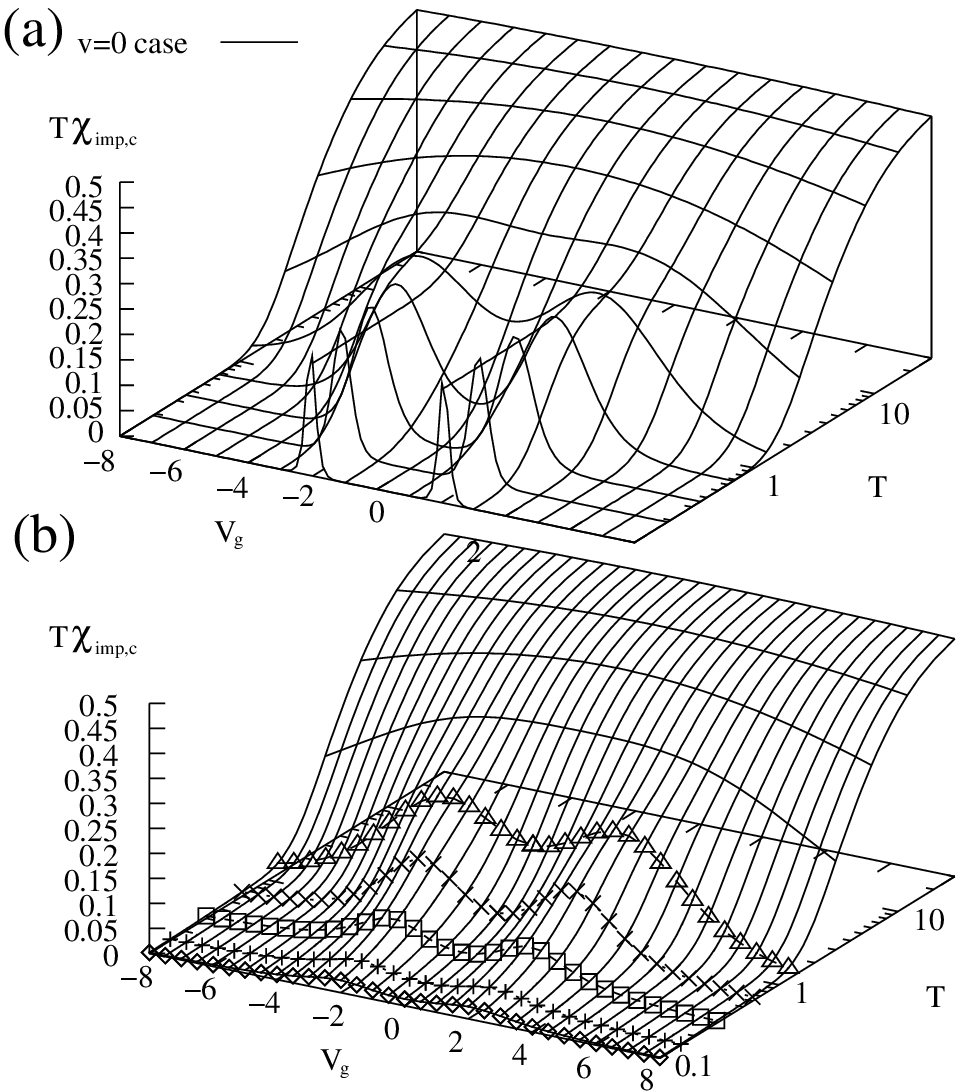}}
\caption{(a) Three dimensional plot of a charge susceptibility $T\chi_{\rm imp,c}$ as a function of $T$ and $V_{\rm g}$ for $U=4$ in the atomic limit $v=0$, 
where $T\chi_{\rm imp,c}={2 \e^{{\epsilon_d +U\over T}} (1+2 \e^{{\epsilon_d \over T}}+\e^{{2\epsilon_d +U\over T}}) \over (1+2\e^{{\epsilon_d +U\over T}}+\e^{{2\epsilon_d +U\over T}})^2}$.
(b) Three dimensional plot of a charge susceptibility $T\chi_{\rm imp,c}$ as a function of $T$ and $V_{\rm g}$ for $U=4$ and $v=1/\sqrt{2}$.}
\label{VxTyTchicv0}
\label{VxTyTchic}
\end{figure}

\begin{figure}[tb]
%\resizebox{8cm}{!}{\includegraphics{fig-new/sz.s4axt0707.chi.ver2.eps}}
\resizebox{8cm}{!}{\includegraphics{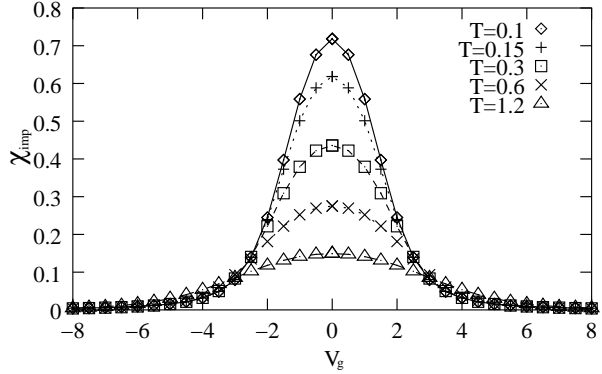}}
\caption{A spin susceptibility $\chi_{\rm imp}$ as a function of $V_{\rm g}$ at various $T$ for $U=4$ and $v=1/\sqrt{2}$. 
}
\label{s4axt0707.chi}
\end{figure}

\begin{figure}[tb]
%\resizebox{8cm}{!}{\includegraphics{fig-new/n.s4axt0707.chic.ver2.eps}}
\resizebox{8cm}{!}{\includegraphics{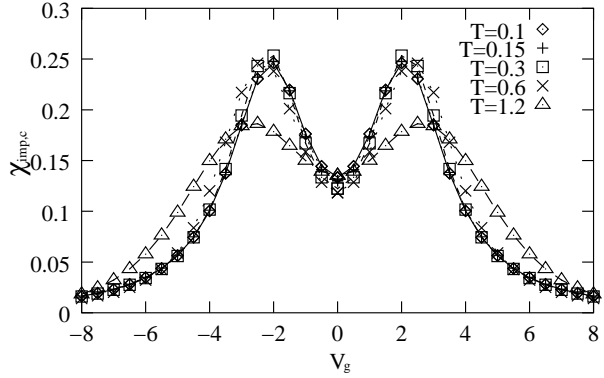}}
\caption{A charge susceptibility $\chi_{\rm imp,c}$ as a function of $V_{\rm g}$ with various $T$ for $U=4$ and $v=1/\sqrt{2}$. 
}
\label{s4axt0707.chic}
\end{figure}

Finally, we discuss low temperature properties of QD by comparing  $\chi_{\rm imp}$ (Fig.~\ref{s4axt0707.chi}),
$\chi_{\rm imp,c}$ (Fig.~\ref{s4axt0707.chic}) and $g$ (Fig.~\ref{conductanceaU4}).
When $|V_{\rm g}|>2$, temperature dependences of these quantities show convergence already at $T=0.1$.
On the other hand, when the number of electrons is odd ($|V_{\rm g}|<2$), 
the fixed point at zero temperature is far from the lowest temperature $T=0.1$ in the present calculations,
as can be seen from the conductance which does not reach the unitarity limit at least in the symmetric case.
This behavior is consistent with the fact that $\chi_{\rm imp}$ has a maximum at the symmetric point (Fig.~\ref{s4axt0707.chi}),
because the inverse of $\chi_{\rm imp}(T=0)$ is proportional to the cross-over temperature when $|V_{\rm g}|<2$, 
i.e. a large $\chi_{\rm imp}$ means a small cross-over temperature.
On the other hand,
the charge susceptibility $\chi_{\rm imp,c}(T=0)$ and the conductance $g(T=0)$ reflect the number of electrons as the function of $\epsilon_d$.
Their behaviors are quite different when the number of electrons is odd.
However,
their basic behavior is easy to understand because the number of electrons changes like a step function with increasing $\epsilon_d$.
The reason why  $\chi_{\rm imp,c}$ is not zero at $V_{\rm g}=0$ in Fig.~\ref{s4axt0707.chic} is that $v^2/U$ is finite and not small.
When $v^2/U$ is small, $n_d$ becomes flatter around the symmetric point and $\chi_{\rm imp,c}$ becomes smaller.

It can be concluded that we can observe the valence-fluctuation regime as visible peaks in the conductance and in the charge susceptibility,
because the valence-fluctuation regime is sensitive to the conductance and also to the charge susceptibility when $V_{\rm g}$ is swept at a fixed temperature.
These peaks give rise to Coulomb oscillation peaks at higher temperatures.
When the strong-coupling regime appears at lower temperatures,
the charge susceptibility and the magnetic susceptibility are saturated,
i.e.
$\chi_{\rm imp,c}$ and $\chi_{\rm imp}$ takes finite values.
Since the cross-over temperature from the valence-fluctuation to the strong-coupling regime is thought to be higher than that at the symmetric point $V_{\rm g}=0$,
Coulomb oscillation peaks in the conductance gradually move to $V_{\rm g}=0$ with decreasing temperature as we see in Fig.~\ref{conductanceaU4}.
At last, the conductance at $V_{\rm g}=0$ should reach the unitarity limit,
because the conductance at zero temperature follows eq.~\ref{eqg_sum}.
However, at low temperatures where the conductance is close to the unitarity limit,
there is the difficulty of the numerical analytic continuation
as we mentioned in the previous section.

\section{Summary}
We have developed a method to calculate thermo-dynamical quantities and the conductance at various parameters simultaneously by using the F$T$-DMRG,
because F$T$-DMRG can use the same result for the bulk part in the calculations of various $v$, $U$ and $\epsilon_d$.
This is the advantage of the F$T$-DMRG; for example, the bulk-part calculation needs a large memory mainly in a Lanczos routine for the quantum transfer matrix, but the impurity-part calculation does not need such a process.
What we should do next is to calculate the average of physical quantities.
Actually we have seen that cross-over behaviors between different regimes are observable not only in the conductance but also in the magnetic and charge susceptibilities by changing impurity parameters.
Especially we can clearly see nice correspondence between the Unitarity limit and the strong-coupling regime and also between the Coulomb oscillation peaks and the valence-fluctuation regime.
We may conclude that F$T$-DMRG is useful method to study the QD.

Next we will discuss some technical aspect of the F$T$-DMRG.
Since F$T$-DMRG is an approach where the numerical error accumulates as temperature is lowered,
we can not use realistic parameters i.e. $v$ and $U$ are much smaller than the band width $t$,
because in the parameter region the Kondo temperature becomes extremely small.
In principle,
we can separate up and down spin electrons in the bulk part calculation.
By doing this, probably one can go to lower temperatures than the present result.
The reason we did not use the algorithm, is that we will use the present program to more general system such as the case that the Coulomb interaction exists in the leads.
The effect of many body interaction in the leads will be reported in a separate publication.

%%% Error of MEM and Pade
In order to obtain reliable results at  low temperatures with this method,
there is another problem:
the numerical analytic continuation.
For example, as we mentioned already raw data of conductance sometimes become larger than the unitarity limit at $T\ll T_{\rm K}$.
The numerical error comes from two reasons;
one is an error of the F$T$-DMRG in a thermal Green's function $G_0(\tau)$,
and the other is an additional error coming from the numerical analytic continuation from $G_0(\tau)$ to $G_0(\epsilon)$.
The numerical errors of $G_0(\tau)$ at $\tau=0,\beta$
correspond to truncation errors, 
because $G_0(\tau=0)$ is the number of electrons at the dot.
The truncation errors are typically $10^{-3} \sim 10^{-4}$ in this calculation.
Since $G(\tau\sim \beta/2)$ becomes smaller with decreasing temperature,
the relative error becomes larger at lower temperatures.
$G(\tau\sim \beta/2)$ has a stronger dependence both on the Trotter number $M$ and on the maximum number of residual bases $m$.
This is the difficulty concerning the numerical calculation of the thermal green function.
In addition to this, the numerical analytic continuation generate additional errors,
which appear even if $G_0(\tau)$ calculated with a high accuracy from the explicit DOS is used~\cite{BeachGM00}.
However, as we mentioned in \S \ref{conductanceU0} concerning the noninteracting case,
this error becomes small if the conductance is away from the unitarity limit.
It means that one can expect a better results for the conductance than the DOS itself,
because even if the DOS has some error as the function of $\omega$, 
it's integrated value like the conductance may show a better behavior.

 %%Boundary edge moments And definition of \chi
 %But, in the calculation with our model at high temperatures near the band width, we can see an additional feature which originates in moments on next sites to the impurity site, which becomes clear at the atomic limit $v=0$.
 %At the atomic limit, these moments are explained as boundary edge moments on a tight binding chain under the open boundary condition.
 %These boundary edge moments will not appear on the Anderson model.
 %To avoid these effects, we use the susceptibility 
 %$\chi_{\rm imp}=\langle s^z_{i=\impsite}\rangle/h$, where $h$ is a small magnetic field used for the numerical differentiation.
 %
\section*{Acknowledgment}
We are grateful to T. Fujii for useful discussions and comments.
This work is supported by Grant-in-Aid for Scientific Research
from Japan Society for the Promotion of Science
and
by New Energy and Industrial Technology Development Organization (NEDO).

\end{document}